\documentclass[11pt]{JHEP3}

\usepackage{amsmath,amsthm,amsfonts,amssymb,citesort,amscd,mathrsfs,graphicx,fontenc,Glebdefs,
}

\usepackage{color}
\let\normalcolor\relax
\usepackage{colordefs}

\author{Marius de Leeuw$^a$\footnote{E-mail: Marius.de.Leeuw@aei.mpg.de, S.J.vanTongeren@uu.nl.}\, and Stijn J. van Tongeren$^b$\\  
$^a$ Max-Planck-Institut f\"ur Gravitationsphysik, Albert-Einstein-Institute\\
\quad Am M\"uhlenberg 1, 14476, Potsdam, Germany\\
$^b$ \it Institute for Theoretical Physics and Spinoza Institute,\\ 
\quad Utrecht University, 3508 TD Utrecht, The Netherlands}

\abstract{Starting with a discussion of the general applicability of the simplified mirror TBA equations to simple deformations of the $\AdS$ superstring, we proceed to study a specific type of orbifold to which the undeformed simplified TBA equations directly apply. We then use this set of equations, as well as L\"uscher's approach, to determine the NLO wrapping correction to the energy of what we call the orbifolded Konishi state, and show that they perfectly agree. In addition we discuss wrapping corrections to the ground state energy of the orbifolded model under consideration.}

\title{Orbifolded Konishi from the Mirror TBA}

\preprint{
AEI-2011-017\\
ITP-UU-11/11\\
SPIN-11/07
}

\begin{document}
\renewcommand{\thefootnote}{\arabic{footnote}}
\setcounter{footnote}{0}

\section{Introduction}

In this paper we discuss aspects of finite size integrability of an
orbifold of the $\AdS$ superstring. Through the AdS/CFT duality
\cite{Maldacena:1997re}\footnote{For a recent review of integrability in the AdS/CFT duality see \cite{Beisert:2010jr}.}, this provides us with means to determine the
anomalous dimensions of certain operators in the corresponding dual,
orbifolded super Yang-Mills theory.

In the setting of the AdS/CFT correspondence, the only tool that is
currently available to non-perturbatively study finite size effects is a
set of equations known as the mirror thermodynamic Bethe ansatz (TBA)
equations. Based on experience in integrable relativistic models
\cite{Zamolodchikov:1989cf}, the idea of applying similar methods to the
AdS/CFT correspondence was put forward in \cite{Ambjorn:2005wa} and explored in detail in \cite{Arutyunov:2007tc}. The main step in deriving the mirror TBA equations is the formulation of the string hypothesis \cite{Takahashi:19721dHubbard}, which has been taken in \cite{Arutyunov:2009zu} by using the mirror version of the Bethe-Yang equations \cite{Beisert:2005fw} for the $\ads$ superstring. This was followed by a derivation of the canonical \cite{Arutyunov:2009ur,Bombardelli:2009ns,Gromov:2009bc} and simplified \cite{Arutyunov:2009ux} TBA equations, describing the ground state of the theory. The associated Y-system was conjectured in \cite{Gromov:2009tv}. The mirror TBA equations have
subsequently been used to analyze the vanishing of the ground state
energy beyond the asymptotic regime \cite{Frolov:2009in}. Furthermore,
these ground state equations can be used to obtain equations for the
excited states, through a contour deformation trick
\cite{Arutyunov:2011uz} inspired by the analytic continuation procedure
of \cite{Dorey:1996re}. Using this trick, the mirror TBA equations have
been used to reproduce perturbative results found through L\"usher's
approach\footnote{The use of L\"usher's approach \cite{Luscher:1985dn}
in the AdS/CFT correspondence was first advocated in
\cite{Ambjorn:2005wa}.} \cite{Bajnok:2009vm,Arutyunov:2010gb,Balog:2010xa}, and to study certain
states in the $\alg{sl}(2)$ sector in considerable detail \cite{Gromov:2009tq,Arutyunov:2009ax,Balog:2010vf}, specifically at intermediate coupling in \cite{Gromov:2009zb,Frolov:2010wt}. The involved analytic properties of the Y-functions \cite{Arutyunov:2009ax,Cavaglia:2010nm,Cavaglia:2011kd,Arutyunov:2011inprogress} are essential in determining these equations.

The methods used to study finite-size corrections in $\mathcal{N}=4$ SYM
have also been applied to closely related theories; orbifolds and
so-called $\beta$-deformed theories. These generalizations require knowledge of the corresponding deformed transfer matrices, which for generic deformed theories can be constructed either by twisting the undeformed transfer matrix \cite{Arutyunov:2010gu}, or by twisting the $S$-matrix \cite{Ahn:2010ws}. Such twisting procedures were first concretely applied to $\beta$-deformed theories\footnote{$\beta$-deformed theories were introduced in the setting of AdS/CFT in \cite{Lunin:2005jy} and further investigated in \cite{Frolov:2005ty,Frolov:2005dj,Beisert:2005if,Frolov:2005iq,Alday:2005ww}.}, at the level of the transfer matrix in \cite{Gromov:2010dy} and subsequently confirmed through the general setup in \cite{Arutyunov:2010gu}, while at the level of the $S$-matrix this was done in \cite{Ahn:2010yv,Ahn:2010ws}. Both approaches proved to be successful as they correctly reproduced wrapping energy corrections that were computed in $\beta$-deformed SYM \cite{Fiamberti:2008sm,Fiamberti:2008sn}. The general methods of twisting also make it possible to study the more general $\gamma$-deformations and orbifold models based on the $\ads$ superstring, as was e.g. done for orbifolds in \cite{Arutyunov:2010gu}.

In the present paper we consider a deformation of the string model obtained by twisting the boundary conditions of certain fields, which means that the mirror TBA equations derived for the $\ads$ string are not directly applicable. As we will argue however, the modifications of the canonical TBA equations corresponding to such generic deformations should take a rather modest form, and a considerable amount of these modifications should disappear completely from the {\it
simplified} TBA equations \cite{Arutyunov:2009ax,Arutyunov:2009ux}, by nature of their derivation from the canonical ones. In fact, concretely we will consider twisted boundary conditions that give us an orbifold for which the simplified TBA equations are not modified at all! The apparent slight loss of information in going from
the canonical to the simplified TBA equations, is in some sense really
only apparent; because the asymptotic solution for such deformed models
has a considerably different analytic structure, and the simplified TBA
equations are still a set of coupled nonlinear integral equations, it
seems that together with the contour deformation trick, they contain
enough information to uniquely fix the Y-functions. In other words, we
argue that whatever the underlying canonical TBA equations of our
orbifold model are, they should imply the same set of {\it simplified}
TBA equations as for the $\ads$ string. Nevertheless, a derivation from
first principles of these and more generically deformed TBA equations, and a general investigation of
the integrable properties of these models would be both important and interesting.

The model we will consider is an orbifold of the five sphere of the
string theory, dual to a special subset of the $\mathbb{Z}_S$ orbifolds
considered in \cite{Beisert:2005he}, which preserves none of the
supersymmetry of the undeformed model. As a specific application, we
will consider a state which in the undeformed case is the $\sl(2)$
descendant of the Konishi state; for simplicity we will call it
orbifolded Konishi. We will describe the analytic properties of the
asymptotic solution, and use it to find the set of simplified TBA
equations for this excited state. These TBA equations can then be used
to find the leading-order (LO) and next-to-leading-order (NLO) finite
size corrections to the energy of this state, giving the perturbative
anomalous dimension of the corresponding dual operator. Due to the
deformation, these so-called wrapping effects start at considerably
lower order in the coupling constant, and so might be quite readily
accessible by direct computations in the gauge theory.

We will show that these finite size corrections are in perfect agreement with those
computed through L\"uscher's perturbative approach \cite{Luscher:1985dn,Bajnok:2008bm}, providing a
consistency check on both approaches. This is the first instance where
the mirror TBA has been used to study a \emph{physical} state in a
deformed theory, and we see that the puzzling disagreement between the
mirror TBA and L\"uscher corrections, described in \cite{Arutyunov:2010gu} for an unphysical magnon, is not present for this physical state.

We will also briefly comment on the ground state of this theory, which has nonzero energy. As a concrete demonstration of this, we compute the LO and NLO wrapping correction to the energy of the ground state from its asymptotic solution. Moreover, we in fact discuss wrapping corrections up to double wrapping order.

This paper is organized as follows. First we give a general discussion of the
effects of twists on the canonical and simplified TBA equations, and argue that the simplified TBA equations for the specific orbifolds we consider are not modified. We then discuss the orbifolded Konishi state and its asymptotic solution in terms of twisted transfer matrices, which is subsequently used to find the corresponding simplified TBA equations. In the process we also find the asymptotic solution for the ground state. Finally we discuss the perturbative LO and NLO wrapping corrections to the energy of the orbifolded Konishi state, as well as the ground state, and show that L\"uscher's perturbative approach and the TBA approach perfectly agree. We will also briefly touch upon more general twist-2 operators in this context along the lines of \cite{Bajnok:2008qj,deLeeuw:2010ed}.

\section{Applicability of the simplified TBA equations}

In this section we will discuss to what extent the simplified TBA equations as derived in \cite{Arutyunov:2009ur,Arutyunov:2009ux} are applicable to integrable deformations of the $\AdS$ superstring. This would be analogous to the wider applicability of the Y-system to excited states, for undeformed as well as deformed models. The Y-system can be derived from the TBA equations \cite{Arutyunov:2009ur}, through the application of an operator with a large kernel. This means that in general there are (many) solutions to the Y-system that are not solutions of the TBA; it is more universal, but it also contains less information. As discussed in \cite{Arutyunov:2011uz}, the TBA equations are universal in the sense that the form of the equations is the same for any state, with the distinction between states coming from modified integration contours combined with distinguishing analytic properties of the Y-functions for different states. These differences only lead to the appearance of driving terms, driving terms which disappear exactly upon application of the operator mentioned above, as required to obtain the Y-system equations.

In a similar fashion, when deforming this model, provided of course that the assumption of quantum integrability holds\footnote{Specifically beyond the asymptotic regime.}, we should expect the TBA equations to be modified\footnote{The authors would like to thank Zoltan Bajnok for enlightening discussions on this topic.}; chemical potentials should enter the equations, which themselves could be coupled in a more involved fashion. It is still possible to imagine however, that the simplified TBA equations for such models would only be modified in very simple ways, or could even be unchanged in certain cases. A good example of this is the Hubbard model in the presence of a magnetic field, which does not affect the simplified TBA equations \cite{Korepin}. The simplified TBA equations are equivalent to the canonical TBA equations, but only upon specification of certain boundary conditions on the Y-functions\footnote{For the Hubbard model for example, these take the form of conditions like $\log Y_M / M \rightarrow B/T$ as $M \rightarrow \infty$, where $B$ and $T$ are the magnetic field the temperature respectively.}; for us the large $J$ asymptotic solution will play this role, see also figure \ref{fig:YMwasymptotics}. Let us consider these ideas in slightly more detail.

\subsection{General deformations}

Twisting boundary conditions on the string theory side corresponds to twisting boundary conditions in the time direction of the mirror model. Hence, in the process of deriving the TBA equations, the equations describing the relation between the particle and hole densities for the different roots should remain unchanged; it is the definition of the free energy for a deformed model which is modified. Twisted boundary conditions of the model in the time direction can schematically be taken into account in the derivation of the TBA equations by introducing a defect operator \cite{Bajnok:2004jd} in the definition of the partition function \cite{Bajnok:2007jg}
\begin{equation}
Z \sim \mbox{Tr}(e^{\beta \mathcal{H}}) \rightarrow \mbox{Tr}(e^{\beta \mathcal{H}}D).
\end{equation}
We will not discuss the exact form of this operator for specific deformations, though it is clear that it should be related to the twisting of the transfer matrices giving the asymptotic solution in a direct and simple manner, whether it be through a twisted $S$-matrix \cite{Ahn:2010ws}, or through what has been called an operational twist \cite{Arutyunov:2010gu}. Were $D$ simply a 'diagonal' operator, it would directly correspond to adding simple chemical potentials for each type of particle; chemical potentials which depend directly on the twist. However, since from the operational point of view, the twists in the left and right sectors of the mirror model are linked, in general this defect operator will most likely couple the two sectors. Equivalently, by twisting the $S$-matrix the splitting between left and right sectors in general disappears, meaning that $D$ should couple the two sectors. Let us emphasize however, that they should naturally decouple again in the undeformed limit.

In short, we should expect chemical potentials to enter the canonical TBA equations, where these chemical potentials should be proportional to the twist, and couple the left and right sectors in a fairly simple manner. Furthermore, for a string configuration or bound state the chemical potential should clearly be $M$ times the chemical potential of the constituents, where $M$ is the length of the configuration\footnote{E.g. for an $M|vw$ string the chemical potential would schematically be $(\mu_w +2 \mu_y) M$.}.

These more involved canonical TBA equations should still reduce to the same Y-system upon application of the appropriate kernels. In the undeformed case, this is done in two steps; we first applies the kernel $(K+1)^{-1}$ (see \cite{Arutyunov:2009ur} for a complete list of kernels), yielding the simplified TBA equations, which is followed by application of $s^{-1}$, giving the Y-system. Now as $s^{-1}$ has a large kernel, it is not surprising that such a large set of different TBA equations reduces to one and the same Y-system\footnote{Note the additional discussion of this point just below.}. The main purpose of this section is to consider to what extent a similar mechanism might take place one step earlier, this time not between ground state and excited state equations, but between deformed and undeformed equations. To do so, let us consider $(K+1)^{-1}$,
\begin{equation}
(K+1)^{-1}_{MN} = \delta_{MN} - s (\delta_{M+1,N} + \delta_{M-1,N}), \,\, {\rm where} \, \, \, s(u) = \frac{g}{4 \cosh{\frac{g \pi u}{2}}} \, .
\end{equation}
This indeed seems to be a nice inverse to $K+1$;
\begin{equation}
(K+1)^{-1}_{MN}(K_{NQ} +\delta_{NQ}) = \delta_{MQ} ,
\end{equation}
where application of this kernel means a sum over repeated indices, and an integration over the rapidity $u$. As shown explicitly in for instance \cite{Arutyunov:2009ur}, this inverse indeed exists on the space of functions which admit a smooth Fourier transform. However, on constant functions, $K+1$ most certainly does not have an inverse, since constant functions of the form
\begin{equation}
c = \alpha N
\end{equation}
are in the kernel of its would-be inverse
\begin{equation}
\label{eq:Kp1invkernel}
(K+1)^{-1}_{MN}(c) = \alpha (M - \tfrac{1}{2}(M + 1 + M - 1))= 0 \,,
\end{equation}
owing to the normalization of $s$, $\int s = 1/2$. It then immediately follows that when this operator is applied to a canonical TBA equation for a $Q$ particle\footnote{Note that in the $\alg{sl}(2)$ grading we do not necessarily even expect a chemical potential for $Q$ particles, e.g. the $\beta$ deformed model.}, or one for a $M|w$ or $M|vw$ string\footnote{The string hypothesis is the same because the Bethe equations of the mirror model are unchanged.}, the chemical potential of such a configuration will drop out, no matter whether the splitting between left and right sectors is manifestly present. This owes simply to the charge of such a configuration under the defect operator, or twist, \textit{cf.} (\ref{eq:Kp1invkernel}). Hence by applying $(K+1)^{-1}$ to the canonical TBA equations for a deformed model, we end up with a set of equations which no longer contain chemical potentials for any string configuration or bound state of fundamental particles, which as just discussed is one of the main differences between the canonical TBA equations for the undeformed model and deformed models.

This conspicuously leaves out the case of $y$-particles. However, as noted in \cite{Arutyunov:2009ur}, for the undeformed model, reality of the free energy requires a relative sign between the chemical potentials of $y$-particles of type $(1)$ and $(2)$, \textit{i.e.} left and right. Provided this feature persists, and we imagine the simplest possible coupling between the left and right sectors, the chemical potentials present there could very naturally cancel in some equations. In fact, since in the undeformed model the Y-system equations for $y$-particles are obtained from their TBA equations by simply applying $s^{-1}$, which on constant functions has only $\pi i$ in its kernel upon exponentiation, if we expects the undeformed Y-system to come out as a result, no chemical potential could have entered the equation in the first place. However, note that already at the level of the undeformed model it was found that a potential chemical potential for $y$-particles will make its appearance in the Y-system \cite{Arutyunov:2009ur}, and that this cannot be removed from all Y-system equations by a simple redefinition of the Y-functions.

As such it is not unreasonable to expect that models obtained by simple deformations require only modest modifications of the simplified TBA equations. In fact, the model we will concretely consider here satisfies exactly the same simplified TBA equations as the undeformed model. Furthermore, these ideas seem to apply quite well also in the case of $\gamma$-deformations \cite{Zoltan:inprogress}.

Now the key point is that these (deformed) simplified TBA equations should contain enough information to find a unique set of TBA equations for a set of Y-functions corresponding to a specific excited state; they are still coupled integral equations, unlike the algebraic Y-system relations. As discussed in \cite{Arutyunov:2011uz}, the excited state TBA equations are found through the contour deformation trick, using the asymptotic solution and its analytic properties\footnote{The contour deformation trick can of course also be applied to find the canonical excited state TBA equations from the ground state ones, found through introduction of a defect operator as indicated above.}. Now the asymptotic solution for these deformed models is readily available, as described clearly in \cite{Arutyunov:2010gu}, and the analytic properties most certainly depend essentially on the deformation. Therefore, together with the simplified TBA equations this should provide enough information to find the set of TBA equations corresponding to such a state. This was already illustrated for the unphysical single magnon state in \cite{Arutyunov:2010gu}, and in the present paper we will clearly illustrate this for the orbifolded Konishi state.

We would like to emphasize that we have not constructed, or proved existence of the canonical TBA equations for a arbitrary deformations. Rather, we have argued that any possible canonical TBA equations might very well be compatible with (simple modifications of) the simplified TBA equations of the \emph{undeformed} model. The existence of the canonical TBA equations for orbifolded models can to some extent be directly inferred, as there are no major leaps required in the assumptions and derivations of \cite{Arutyunov:2007tc,Arutyunov:2009zu,Arutyunov:2009ur,Arutyunov:2009ux}, but it would be interesting to see their detailed form. Moreover, for $\beta$-deformed theories the story changes considerably more \cite{Frolov:2005ty}. Hence, while in both cases the canonical TBA equations might be significantly more involved than necessary for practical purposes (the simplified TBA equations), it would still be of considerable interest to see a derivation of them.

Now let us discuss the model we will consider in the rest of the paper, and how the above ideas give exactly the undeformed simplified TBA equations for this specific deformation.

\subsection{Orbifolded Konishi}

In this paper we consider a particular $\mathbb{Z}_S$ orbifold of the regular $\AdS$ string theory, which falls general class of orbifolds considered in \cite{Beisert:2005he}. The orbifold we consider was first considered in \cite{Arutyunov:2010gu} in the TBA approach, and corresponds to twisting the boundary conditions of the bosons $y_{a}$ and $y_{\dot{a}}$, for $a=1,2$ and $\dot{a}=\dot{1},\dot{2}$, as
\bea\nonumber
\left(\begin{array}{c} y_{1}(2\pi) \\ y_{2}(2\pi)
\end{array} \right)=\left(\begin{array}{cc}
e^{i\alpha_{\ell}} & 0 \\ 0 & e^{-i\alpha_{\ell}}
\end{array} \right)\left(\begin{array}{c}
y_{1}(0) \\ y_{2}(0)
\end{array} \right)\,
, ~~~~~~ \left(\begin{array}{c} y_{\dot{1}}(2\pi) \\
y_{\dot{2}}(2\pi)
\end{array} \right)=\left(\begin{array}{cc}
e^{i\alpha_r} & 0 \\ 0 & e^{-i\alpha_r}
\end{array} \right)\left(\begin{array}{c}
y_{\dot{1}}(0) \\ y_{\dot{2}}(0)
\end{array} \right).
\eea

The twists above will lead to a modification of the boundary conditions for two of the four fields $Y_{a\dot{a}}$, which together with the two fields contained in the unaffected complex $Z$ field parametrize the five-sphere. Moreover, by construction the physical fermions $\theta_{a\dot{\alpha}}=y_a\theta_{\dot{\alpha}}$ and $\eta_{\dot{a}\alpha }=\theta_{\alpha}y_{\dot{a}}$ inherit the twisted periodicity conditions. To obtain a nice orbifold we take $\alpha_{\ell}=\pm\alpha_r$, with $\alpha_{\ell}=\tfrac{2 \pi n}{S}$; the choice of sign does not matter as the two resulting theories are essentially equivalent. In fact, for states without auxiliary excitations the transfer matrices is manifestly symmetric in $\alpha$.

The defect operator $D$ introduced above should take a very simple form for this type of orbifold. Effectively it should simply add the appropriate chemical potentials to the canonical TBA equations. In fact, since we only twist the boundary conditions for the $w$ excitations and not the ones for $y$-particles (they correspond to fundamental fermions), only chemical potentials for $w$ strings should appear. However, as discussed above these will immediately drop out from the simplified equations. \emph{This means we can use the undeformed simplified TBA equations of \cite{Arutyunov:2009ux} to study this type of orbifolded models.}

Within this type of orbifolded theories we study the $\alg{sl}(2)$ descendant of the Konishi state, which we call the orbifolded Konishi state, as indicated in the introduction. This state has $J=2$ and excitation numbers $K^{I} = M = 2$ and $K^{II} = K^{III} =0$, meaning that is has only two fundamental excitations, and no auxiliary ones. Its AdS/CFT correspondent operator is schematically of the form $\mbox{Tr}(\gamma^n Z D^2 Z)$, where $\gamma$ is the appropriate generating element of $\mathbb{Z}_S$. It has length two, the two Z's correspond to $J=2$ and the covariant derivatives correspond to the fundamental excitations on the string side. The ground state in the $\alg{sl}(2)$ sector depends on the length and is given by $\mbox{Tr}(\gamma^n Z^J)$.

The main ingredient in either the TBA or L\"uscher's corrections is the asymptotic solution, constructed through the use of left and right transfer matrices, $T^{l,r}_{Q,1}$. For states without auxiliary excitations, and an arbitrary twist of the type above (meaning $T^{l}_{Q,1} = T^{r}_{Q,1} = T_{Q,1}$), these are given by
\bea\label{TS}
T_{Q,1}(v\,|\,\vec{u})&=&1+\prod_{i=1}^{M} \frac{(x^--x^-_i)(1-x^-
x^+_i)}{(x^+-x^-_i)(1-x^+
x^+_i)}\frac{x^+}{x^-}\\
&&\hspace{-1.5cm}-2\cos\alpha\sum_{k=0}^{Q-1}\prod_{i=1}^{M}
\frac{x^+-x^+_i}{x^+-x^-_i}\sqrt{\frac{x^-_i}{x^+_i}}
\left[1-\frac{\frac{2ik}{g}}{v-u_i+\frac{i}{g}(Q-1)}\right]+\sum_{m=\pm}
\sum_{k=1}^{Q-1}\prod_{i=1}^{M}\lambda_m(v,u_i,k)\, . \nonumber
\eea Definitions of various quantities entering the last formula
can be found in appendix \ref{ap:twistedtransfermatrix} (see also \cite{Arutyunov:2009iq}); $\alpha$ is the parameter describing the twist of the bosonic eigenvalues, note how it enters symmetrically in the transfer matrix.

In addition to this information, in order to find the simplified TBA equations for this excited state from the ground state ones we need the solution of the Bethe-Yang equations for the momenta of the two excitations. Fortunately for the diagonal types of twist used above, in the $\alg{sl}(2)$ grading the main Bethe-Yang equation is not modified so that it has the same solution as the true Konishi state, found numerically in \cite{Arutyunov:2009ax}.

In the next section we will discuss how the asymptotic solution for orbifolded Konishi allows us to find a working set of simplified TBA equations. We will then calculate the NLO finite size correction to the energy and show that L\"uscher's perturbative approach is in perfect agreement with the TBA approach as a check on the validity of both approaches.

\section{The TBA equations for orbifolded Konishi}

In order to obtain TBA equations for an excited state, following \cite{Arutyunov:2009ax}, we employ the contour deformation trick. A clear overview of this whole approach has most recently been given in \cite{Arutyunov:2011uz}. In short we assume that the ground state and excited state TBA
equations differ only by the choice of the integration contours,
and upon deforming the integration contours of the excited state
TBA equations to the ground state ones, we pick up contributions
of extra singularities, leading to the appearance of new driving
terms in the excited state TBA equations. Below we
discuss the analytic properties and related integration contours for the orbifolded Konishi state in detail, followed by the resulting TBA equations.

\subsection{Analytic properties}

\begin{figure}
\begin{center}
 \includegraphics[width=5in]{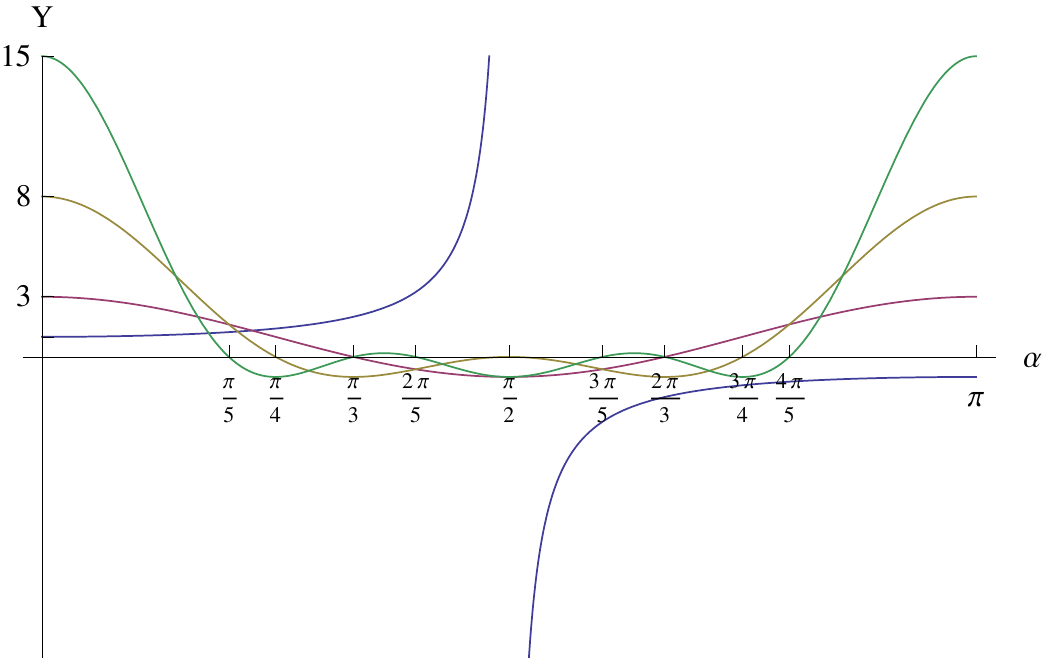}
\end{center}
\caption{The asymptotics of $Y_\pm$ (blue), $Y_{1|w}$ (red), $Y_{2|w}$ (yellow), $Y_{3|w}$(green) as a function of the twist.}
\label{fig:YMwasymptotics}
\end{figure}

Naturally, the analytic properties of the Y-functions depend heavily on the twist, while only their finer details depend on the coupling constant $g$. Most importantly, while the asymptotics of $Y_{M|vw}$ on the mirror line  are $M(M+2)$ as should be expected,
\begin{equation}
\lim_{v\rightarrow \infty} Y_{M|vw}(v) = M(M+2),
\end{equation}
the asymptotics of $Y_{M|w}$ and $Y_\pm$, depend on the twist, and indicate interesting behaviour; of real interest is the fact that the asymptote can be both positive and negative. The asymptotic values of both $Y_\pm$ are given by
\begin{equation}
\lim_{v\rightarrow \infty} Y_{\pm}(v) = \sec \alpha,
\end{equation}
while those of $Y_{M|w}$ are given by
\begin{equation}
\lim_{v\rightarrow \infty} Y_{M|w}(v) = M + 2\sum_{Q=0}^{M}(M-Q)\cos (2Q+2)\alpha .
\end{equation}
These results can be readily derived by realizing that in this limit the expression for the transfer matrix reduces to the character of the representation of the group element corresponding to the twist. Moreover, these constants are the Y-functions we should expect to find as the asymptotic ground state solution. Indeed, it is easily verified that the above expressions satisfy the ground state equations for $w$ and $vw$-strings, and $y$ and $Q$-particles. To go beyond the asymptotic regime requires considerably more work of course. Because there is no more supersymmetry to protect the ground state energy from getting finite size corrections, we will present the leading order wrapping correction to the ground state energy in section \ref{subsec:LO}, which is readily found from this solution. In fact it would be interesting to see an analysis of the ground state TBA equations along the lines of \cite{Frolov:2009in} and \cite{Frolov:2010wt} for this orbifold. In figure \ref{fig:YMwasymptotics} we have plotted the asymptotics of $Y_-$, $Y_{1|w}$, $Y_{2|w}$, and $Y_{3|w}$.

It is interesting to note that the asymptotics in the zero twist limit smoothly reduce to the known values of $1$ and $M(M+2)$ respectively, but also that this almost happens when the twist parameter is equal to $\pi$,
\begin{align}
\lim_{v\rightarrow \infty} Y^{\alpha = 0,\pi}_{M|w}(v)&  = M(M+2),\\
\lim_{v\rightarrow \infty} Y^{\alpha = 0}_{\pm}(v)& = -\lim_{v\rightarrow \infty} Y^{\alpha = \pi}_{\pm}(v) = 1.
\end{align}
This is of course a very particular value, but still we clearly see that the asymptotic solution is really different from the untwisted case.

The fact that the asymptotic values of the $Y_{M|w}$-functions can change from positive to negative means that these functions must have roots on the real mirror line for certain values of the twist and coupling; roots which are not present for the true Konishi state \cite{Arutyunov:2009ax}. The rest of this section will be dedicated to a concrete discussion of the analytic properties of the Y-functions.

Let us start by recalling the construction of the asymptotic Y-functions in terms of the transfer matrices\footnote{The general construction of the Y-functions in terms of transfer matrices is based on the underlying symmetry group of the model\cite{Kuniba:1993cn,Tsuboi}; for the string sigma model asymptotic Y-functions were presented in \cite{Gromov:2009tv}. In fact, this solution can be derived from the AdS/CFT Y-system together with the Bajnok-Janik formula \cite{Bajnok:2008bm}, see \cite{Arutyunov:2009ax}.},
\begin{align}
Y_{M|w} & = \frac{T_{1,M} T_{1,M+2}}{T_{2,M+1}} \, ,  \, \, \, \, \, Y_{-} = -\frac{T_{2,1}}{T_{1,2}} \, ,  \, \, \,\, \,  Y_{+} = - \frac{T_{2,3}T_{2,1}}{T_{1,2}T_{3,2}} \notag \, ,\\
Y_{M|vw}& = \frac{T_{M,1} T_{M+2,1}}{T_{M+1,2}}  = \frac{T_{M,1} T_{M+2,1}}{T_{M+1,1}^- T_{M+1,1}^+ - T_{M,1} T_{M+2,1}}.\label{eq:YmvwinT} 
\end{align}
Here $T_{M,Q}^\pm$ denotes $T_{M,Q}$ with its argument shifted by $\pm i/g$. Roots of $Y_{M|w}$ arise from the roots of $T_{1,M}$ and $T_{1,M+2}$, while roots of $Y_{M|vw}$ arise from the roots of $T_{M,1}$ and $T_{M+2,1}$. For $Y_{M|vw}$ it is immediately clear that its roots shifted by $\pm i/g$ give roots of $1+Y_{M-1|vw}$ and $1+Y_{M+1|vw}$,
\begin{equation}
1+Y_{M|vw}(r_{\scriptscriptstyle M}^\pm) = 0, \, \,{\rm where} \, \, \, T_{M,1} (r_{\scriptscriptstyle M-1}) = 0 \, .
\end{equation}
While not obvious from their expression, a similar relation holds for $Y_{M|w}$,
\begin{equation}
1+Y_{M|w}(\rho_{\scriptscriptstyle M}^\pm) = 0, \, \,{\rm where} \, \, \, T_{1,M} (\rho_{\scriptscriptstyle M-1}) = 0.
\end{equation}
The analytic properties for the orbifolded Konishi state are quite involved, in the sense that a large number of roots appear and play a role in the TBA, with many of them displaying critical behaviour of a nature that depends on the specific value of the twist. In the main text we will refrain from discussing these technical complications and focus on the TBA equations at small coupling; these are the equations that are presently of most value. A short discussion on the finer details of the analytic properties can be found in appendix \ref{ap:criticality}. 

\begin{figure}
\begin{center}
 \includegraphics[width=5in]{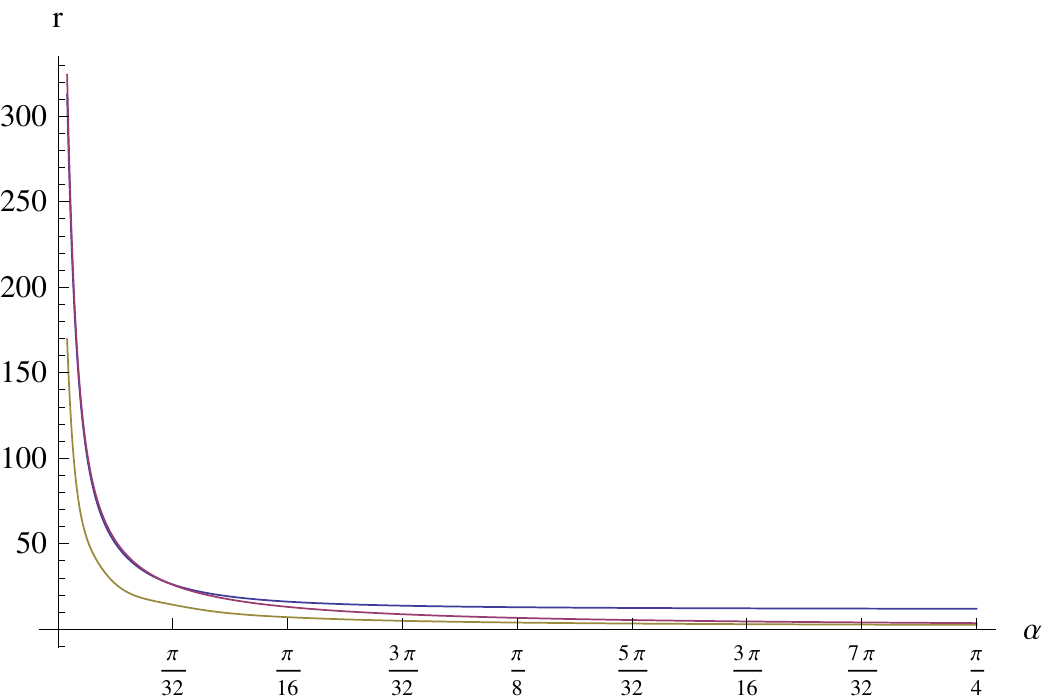}
\end{center}
\caption{The movement of the real root $\mathbf{r}_1$ of $1+Y_{1|vw}(v-i/g)$ with the twist, for $g=\tfrac{1}{10}$ (blue), $g=1$ (red), and $g=10$ (yellow). The root asymptotes to the value of $2$ as the coupling is increased.}
\label{fig:Y1vwroottwistbehaviour}
\end{figure}

At small coupling, the $Y_{M|vw}$-functions have four roots, which we will denote by $\pm \mathbf{r}_{\scriptscriptstyle M\pm1}$\footnote{As we are discussing the asymptotic solution, these roots of course correspond to roots of the transfer matrices $T_{M,1}$ and $T_{M+2,1}$,  which we denote by $\pm \mathbf{r}_{\scriptscriptstyle M-1}$, and $\pm \mathbf{r}_{\scriptscriptstyle M+1}$ respectively.}. These roots were not observed for the true Konishi state, which agrees with our findings, since these roots move away towards infinity as the twist is removed, see figure \ref{fig:Y1vwroottwistbehaviour}. As these roots are real, their shifted counterparts, the roots of $1+Y_{M\pm1|vw}$, will always appear in the TBA for orbifolded Konishi.
Just as in the Konishi case, $Y_+$ has poles at $u_i^\pm$, which lead to the appearance of additional driving terms in the TBA equations. While not discussing criticality here, let us briefly note that we reproduce the critical behaviour found in \cite{Arutyunov:2009ax}, coming about as the smooth $\alpha \rightarrow 0 $ limit of the general behaviour, see appendix \ref{ap:criticality} for more details.

To summarize, at small coupling the following points are important in the simplified TBA equations,
\begin{equation}
Y_+(u_i^\pm)  = \infty\, , \hspace{10pt}Y_{M|vw}(\mathbf{r}_M^\pm) = Y_{M|vw}(-\mathbf{r}_M^\pm) = -1 \, .
\end{equation}

\subsection{The simplified TBA equations}

Following the contour deformation trick, in order for the asymptotic solution to be a solution, we take the ground state TBA equations of \cite{Arutyunov:2009ur,Arutyunov:2009ux} and define the integration contour such that it goes slightly below the line $-i/g$, \textit{i.e.} such that it encloses the poles of $Y_+$ at $u_i^-$ and the roots of $1+Y_{M|vw}$ at $\pm \mathbf{r}_M^-$ between itself and the real line. By taking the integration contour back to the real line, we find the appropriate driving terms and obtain the TBA equations. Note that the integration kernels and $S$-matrices which enter in the equations below have been defined and are completely listed in \cite{Arutyunov:2009ax}.
\vspace{10pt}\\
\bigskip
 \noindent
$\bullet$\ $M|w$-strings; $\ M\ge 1\ $, $Y_{0|w}=0$
\begin{equation}
\log Y_{M|w} =  \log(1 +  Y_{M-1|w})(1 +
Y_{M+1|w})\star s
 + \delta_{M1}\, \log{1-{1\ov Y_-}\ov 1-{1\ov Y_+} }\hstar s \nonumber,~~~~~
\end{equation}
These equations are not modified at small coupling.\vspace{10pt}\\
\bigskip
 \noindent
$\bullet$\ $M|vw$-strings; $\ M\ge 1\ $, $Y_{0|vw}=0$ 
\begin{align}
\log Y_{M|vw}(v) = & - \log(1 +  Y_{M+1})\star s +
\log(1 +  Y_{M-1|vw} )(1 +  Y_{M+1|vw})\star
s\\
& - \log{S(\mathbf{r}^{-}_{M\pm1} - v)S(-\mathbf{r}^{-}_{M\pm1} - v)} \nonumber\\
&  + \delta_{M1} ( \log{1-Y_-\ov 1-Y_+}\hstar s -  \log{S(\pm u_1^- - v)})\,\nonumber.
\end{align}
Here the driving terms arise from the poles of $Y_+$ and the roots of $1 +  Y_{Q|vw}$; an implicit sum over $\pm$ is implicitly assumed here and below.\vspace{10pt}\\
\bigskip
 \noindent
$\bullet$\   $y$-particles 
\begin{align}
\log {Y_+\ov Y_-}(v) = \, &  \log(1 +  Y_{Q})\star K_{Qy} - \log S_{1_*y}(\pm u_1 ,v)  \,,\\
\log {Y_- Y_+}(v) = \, &  2 \log{1 +  Y_{1|vw} \ov 1 +  Y_{1|w} }\star s - \log\left(1+Y_Q \right)\star K_Q + 2 \log(1 +Y_{Q})\star K_{xv}^{Q1} \star s \\
& - 2\log{S(\mathbf{r}_1^- - v)S(-\mathbf{r}_1^- - v)}-  \log {\big(S_{xv}^{1_*1}\big)^2\ov S_2}\star s(\pm u_1,v) \nonumber
\end{align}
In both equations we get contributions from the exact Bethe equation $Y_1(u_{*i}) = -1$, where the star signifies analytic continuation of the rapidity to the string region\footnote{Similarly, $S_{1_*y}(u_{j},v) \equiv S_{1y}(z_{*j},v)$ is shorthand notation for the S-matrix with the first and second arguments in the string and mirror regions, respectively. The same convention is used for other kernels and S-matrices.}. In the second equation the roots of $1 +  Y_{1|vw}$ also contribute.  For the contribution from the exact Bethe equation we have used the following notation
\bea\nonumber
 &&\log  {\big(S_{xv}^{1_*1}\big)^2\ov S_2}\star s(u,v) \equiv  \int_{-\infty}^\infty\, dt\, \log  {S_{xv}^{1_*1}(u,t)^2\ov S_2(u-t)}\, s(t-v)
 \,.~~~~~
\eea
The contribution then follows from the identity
\begin{equation}
\log{S_1(u_j-v)} - 2 \log{S_{xv}^{1_* 1}}\star s(u_j,v) = -\log{\frac{(S_{xv}^{1_* 1})^2}{S_2}}\star s(u_j,v) ,
\end{equation}
\noindent valid for real $u_j$.\vspace{10pt}\\
\bigskip
 \noindent
$\bullet$\ $Q$-particles
\begin{align}
\log Y_Q(v) = & - L_{\scriptscriptstyle TBA}\, \tH_{Q} + \log \left(1+Y_{Q'} \right) \star \(K_{\sl(2)}^{Q'Q} + 2 \, s \star K^{Q'-1,Q}_{vwx} \)  - \log S_{\sl(2)}^{1_*Q}(\pm u_1,v)\nonumber\\
&  +  2 \log \(1 + Y_{1|vw}\) \star s \hstar K_{yQ} + 2 \, \log \(1 + Y_{Q-1|vw}\) \star s\nonumber\\
&  - 2  \log{1-Y_-\ov 1-Y_+} \hstar s \star K^{1Q}_{vwx} +  \log {1- \frac{1}{Y_-} \ov 1-\frac{1}{Y_+} } \hstar K_{Q}  +  \log \big(1-\frac{1}{Y_-}\big)\big( 1 - \frac{1}{Y_+} \big) \hstar K_{yQ}  \nonumber\\
& -2 \log{S}\hstar K_{yQ} (\pm \mathbf{r}_1^-,v) -2 \log{S}(\pm \mathbf{r}_{Q-1}^-,v)\nonumber\\
& + 2 \log{S}\star_{p.v} K^{1Q}_{vwx} (\pm u_1^-,v) - \log{S^{1Q}_{vwx}} (\pm u_1,v) \, .\label{eq:hybrid}
\end{align}
\noindent Note that here we use the so-called hybrid form of the TBA equations for $Q$-particles as they are more suitable for numerical computations; see \cite{Arutyunov:2009ax} for their derivation.

\smallskip

In the above, $K^{0,Q}_{vwx}=0$ and $Y_{0|vw}=0$, which implies that the $\log S (\pm \mathbf{r}_{0}^-,v)$ terms are also not
present. The principal value prescriptions are required due to the pole of $S(v)$ at $v=-i/g$.

Evaluating (\ref{eq:hybrid}) on the asymptotic solution, for $Q=1$ the equation becomes
\begin{align}
\label{eq:hybridQ=1}
\log{T_{1,1}^2} = & - (L_{\scriptscriptstyle TBA}-J)\, \tH_{1}+  2 \log \(1 + Y_{1|vw}\) \star s \hstar K_{y1}\\
&  - 2  \log{1-Y_-\ov 1-Y_+} \hstar s \star K^{11}_{vwx} +  \log {1- \frac{1}{Y_-} \ov 1-\frac{1}{Y_+} } \hstar K_{1}  +  \log \big(1-\frac{1}{Y_-}\big)\big( 1 - \frac{1}{Y_+} \big) \hstar K_{y1}  \nonumber\\
& -2 \log{S}\hstar K_{y1} (\pm \mathbf{r}_1^-,v) + 2 \log{S}\star_{p.v} K^{11}_{vwx} (\pm u_1^-,v) - \log{S^{11}_{vwx}} (\pm u_1,v) \, .\nonumber
\end{align}
This equation is satisfied, provided that the relation between $L_{\scriptscriptstyle TBA}$ and $J$ is
\begin{equation}
L_{\scriptscriptstyle TBA} = J.
\end{equation}
This is different from the relationship found for the undeformed model where it was found that $L_{\scriptscriptstyle TBA} = J+2$ \cite{Arutyunov:2009ax}, meaning that the relationship between $L_{\scriptscriptstyle TBA}$ and $J$ changes discontinuously as the twist is removed. This has a natural explanation, discussed just below. However also practically this is perhaps not too surprising, as only for strictly zero twist the additional roots of $Y_{M|vw}$ disappear, meaning that the analytic structure is qualitatively different. The continuity of the full equation is discussed in appendix \ref{ap:criticality}.

The relationship between $L_{\scriptscriptstyle TBA}$ and $J$ was nicely explained in \cite{Arutyunov:2011uz}, where it was argued that a single set of TBA equations describes an entire superconformal multiplet of states. With this understanding, $L_{\scriptscriptstyle TBA}$ is the maximal $J$ charge occurring in the multiplet, giving four for the Konishi multiplet as needed. The orbifold which we consider breaks the supersymmetry of the model completely, meaning that the multiplet structure is gone, and so we should expect to find $L_{\scriptscriptstyle TBA} = J$, as we do. The supersymmetry is enhanced at exactly zero twist, which explains the discontinuous relationship between $L_{\scriptscriptstyle TBA}$ and $J$.

\subsection{The exact Bethe equation}

Beyond the asymptotic regime, the rapidity of a particle is determined from the exact Bethe equation $Y_1(u_{*i}) = -1$, replacing the Bethe-Yang equation. The exact Bethe equation is but an analytic continuation away from the TBA equation for a $Q=1$ particle, and continuing the hybrid equation for $Q=1$ to the string region immediately gives,
\begin{align}
\pi i(2n+1) = &  i L_{\scriptscriptstyle TBA}\, p_k - \log \left(1+Y_{1} \right) \star \(K_{\sl(2)}^{11_*} - 2 \, s \star K^{Q-1,1_*}_{vwx} \) - \, \log S_{\sl(2)}^{1_*1_*}(\pm u_1,u_k)\nonumber\\
&+ 2 \log \(1 + Y_{1|vw}\) \star \( s \hstar
K_{y1_*} + \ts\) - 2  \log{1-Y_-\ov 1-Y_+} \hstar s
\star_{p.v.} K^{11_*}_{vwx} \nonumber \\
& -  \log{1-Y_-\ov1-Y_+} \hstar s + \log {1- \frac{1}{Y_-} \ov 1-
\frac{1}{Y_+} } \hstar K_{1} +  \log \big(1-
\frac{1}{Y_-}\big)\big( 1- \frac{1}{Y_+} \big) \hstar
K_{y1_*}\nonumber\\
& -2 \log{S}\hstar K_{y1_*} (\pm \mathbf{r}_1^-,u_k) + 2 \log{S}(\pm \mathbf{r}_1-u_k) \label{eq:ExactBethe}\\
& + 2 \, \log {\rm Res}(S)\star K^{11_*}_{vwx}(\pm u_1^-,u_k) - 2 \sum_j \log{(u_j - u_k - \tfrac{2i}{g})\,\frac{x_j^- -\tfrac{1}{x_k^-}}{x_j^- - x_k^+}} \, .\nonumber
\end{align}
In the above we use the notation
\bea
&&\log {\rm Res}(S)\star K^{11_*}_{vwx} (u^-,v) = \int_{-\infty}^{+\infty}{\rm d}t\,\log\Big[S(u^- -t)(t-u)\Big] K_{vwx}^{11*}(t,v)\,,~~~\notag\\
&&\ts(u)=s(u^-)\notag\,, \eea
the momentum of the magnon is $p = i
\tH_{Q}(z_{*})=-i\log{x_s(u+{i\ov g})\ov x_s(u-{i\ov g})}$, and
the second argument in all kernels in (\ref{eq:ExactBethe}) is the
Bethe root $u_k$.

Below we will be considering the leading order correction to the particle rapidity due to finite size effects, which means we will be expanding the above equation about the asymptotic solution. In other words, we will be considering $\delta \mathcal{R}_k$, where $\mathcal{R}_k$ is given by,

\begin{align}
\mathcal{R}_k \equiv &  \, 2 \log \(1 + Y_{1|vw}\) \star \( s \hstar
K_{y1_*} + \ts\) - 2  \log{1-Y_-\ov 1-Y_+} \hstar s
\star_{p.v.} K^{11_*}_{vwx} \nonumber \\ 
& -  \log{1-Y_-\ov
1-Y_+} \hstar s + \log {1- \frac{1}{Y_-} \ov 1-
\frac{1}{Y_+} } \hstar K_{1} + \log \big(1-
\frac{1}{Y_-}\big)\big( 1- \frac{1}{Y_+} \big) \hstar
K_{y1_*}\nonumber\\
& -2 \log{S}\hstar K_{y1_*} (\pm \mathbf{r}_1^-,u_k) + 2 \log{S}(\pm \mathbf{r}_1-u_k)\label{eq:Rk} \\
& + 2 \, \log {\rm Res}(S)\star K^{11_*}_{vwx}(\pm u_1^-,u_k) - 2 \sum_j \log{(u_j - u_k - \tfrac{2i}{g})\,\frac{x_j^- -\tfrac{1}{x_k^-}}{x_j^- - x_k^+}} = 0 \, , \nonumber
\end{align}
\noindent which are the terms in the exact Bethe equation that do not cancel by default when evaluated on the asymptotic solution.

\section{Wrapping corrections}

In this section we consider the leading order correction to the energy of the orbifolded Konishi state, as well as the ground state. We will also briefly touch upon generic twist-2 states and present their leading order wrapping correction. Subsequently we will focus on the next-to-leading (NLO) correction of the orbifolded Konishi operator. We will first consider the perturbative L\"uscher approach and compare this against the corrections coming from the TBA for the orbifolded Konishi state. 

\subsection{Leading order}
\label{subsec:LO}

The Y-functions are asymptotically given by the generalized L\"uscher's formula \cite{Bajnok:2008bm}
\begin{align}
Y^{\circ}_Q(v) = e^{-J\tilde{\mathcal{E}}_Q(v)}T^{l}(v|\vec{u})T^{r}(v|\vec{u})\prod_i S^{Q1_*}_{\alg{sl}(2)}(v,u_i).
\end{align}
Here $\tilde{\mathcal{E}}_Q(v)$ is the energy of a mirror $Q$-particle, $S^{Q1_*}_{\alg{sl}(2)}(v,u_i)$ denotes the S-matrix with arguments in the mirror ($v$) and string regions ($u_i$) and finally $T^{l,r}$ are the left and right twisted transfer matrices, defined in (\ref{TS}), with $\alpha_{l} = -\alpha_{r} = 2n\pi/S$.

In this section we will study the leading order wrapping correction and correspondingly only evaluate our Y-function to lowest order in $g$. We will denote also this lowest order Y-function simply by $Y^{\circ}_Q$, leaving the $g$-expansion implicit. It turns out that the wrapping correction starts at the two loop level. More precisely, the leading order wrapping correction to the energy ($E_{LO}$) is given by
\begin{align}
E_{LO} = -\frac{1}{2\pi}\sum_{Q=1}^{\infty}\int dv \frac{d\tilde{p}}{dv} Y^{\circ}_{Q}(v).
\end{align}
where, for the orbifolded Konishi operator, the asymptotic Y-function is given by
\begin{align}
Y^{\circ}_Q(v) = \frac{256}{81}g^4\sin^4\frac{\alpha}{2}\frac{Q^2}{(Q^2+v^2)^2}\frac{(3(v^2-\frac{1}{3})-Q^2+1)^2}{
f^+_+f^-_+f^+_-f^-_-}, \quad
f^{\pm}_{\pm}=(Q\pm 1)^2+(v \pm \frac{1}{\sqrt{3}})^2 \, ,
\end{align}
while for the ground state, the asymptotic Y-function is given by,
\begin{equation}
Y^{\circ\scriptscriptstyle{gs}}_Q(v) = 16g^{2J} \sin^4\frac{\alpha}{2}\frac{Q^2}{(v^2+Q^2)^J}.
\end{equation}

Integrating and summing is straightforward and yields the following wrapping correction for orbifolded Konishi, \emph{cf.} \cite{Arutyunov:2010gu}
\begin{align}
E_{LO} = -\frac{g^4}{3}\sin^4\frac{\alpha}{2} \, .
\end{align}
Similarly, for the ground state we find for $J>1$
\begin{equation}\label{eqn;GroundStateLO}
E^{\scriptscriptstyle{gs}}_{LO} = - g^{2J} \frac{8 \Gamma(J-\tfrac{1}{2})\zeta(2 J-3)}{
 \sqrt{\pi} \Gamma(J)} \sin^4\frac{\alpha}{2} \, .
\end{equation}
This is a concrete demonstration of the nonvanishing nature of the ground state energy in the orbifolded model under consideration, as mentioned before. Also, from this result we immediately see that the wrapping correction vanishes in the undeformed ($\alpha \rightarrow 0$) limit, as it should. The wrapping correction in the case $J=1$ is divergent, even though (\ref{eqn;GroundStateLO}) is actually finite at this point. 

Note that the ground state energy diverges logarithmically for $J=2$. It is not clear to us what the origin of this divergence is. Possible causes could be that the two point function of this operator at $J=2$ has special behaviour, {\it e.g.} it could diverge differently than other two point functions. We might also speculate that this theory is not conformal and this is an indication that the two point function is no longer of the form $|x-y|^{-\Delta}$. It would be very interesting to investigate this point from a field theoretic point of view, by considering for example the $\mathbb{Z}_2$ orbifolded theory. Finally, we would like to note that a similar issue was encountered in the undeformed theory \cite{Frolov:2009in}.

Interestingly, by using the twisted transfer matrix from $\beta$-deformed theory from \cite{deLeeuw:2010ed}, we can extend the result for the orbifolded Konishi operator to generic orbifolded twist-2 operators. For a twist-2 operator with spin $M$, the wrapping correction is found to be
\begin{align}
E_{LO} = -  \frac{2 g^4}{M(M+1)} \sin^4\frac{\alpha}{2}.
\end{align}
This result clearly respects reciprocity, and the analytic continuation around $M=-1$ corresponding to BFKL is also completely trivial in this case. 

From the principle of maximum transcendentality we would expect the above result to be of transcendentality degree three. However, since $M^{-1}\sim S_1(M)-S_1(M-1)$ the above seems to be of degree two at most. Nevertheless, as was remarked in \cite{deLeeuw:2010ed}, writing
\begin{align}
E_{LO} = 2 g^4 \sin^4\frac{\alpha}{2}\left[\frac{1}{M+1}-\frac{1}{M}\right],
\end{align}
and assigning degree one to $\sin^2\frac{\alpha}{2}$ gives the expected result of degree three. This might indicate that this is the correct way to interpret the degree of transcendentality in these theories. 

\subsection{Next-to-leading order}

As with the undeformed Konishi state, there are two ways to compute the next-to-leading order correction; the first approach is based on L\"uscher corrections and the second uses the TBA formalism. Both approaches were shown to agree for $\mathcal{N}=4$ SYM \cite{Arutyunov:2010gb,Balog:2010xa,Balog:2010vf}. Here we will similarly use both approaches to compute the NLO correction and will show the necessary agreement between the two. For the groundstate both approaches immediately agree at next-to-leading-order, as there are no excitations.

\subsubsection*{L\"uscher's approach}

There are two different types of terms that contribute to the NLO wrapping correction \cite{Bajnok:2009vm}. First, there is the expansion of the asymptotic Y-function which can be straightforwardly computed from the explicit expression of the transfer matrix, for details see e.g. \cite{Bajnok:2009vm,Arutyunov:2010gu}.  

The other contribution comes from the fact that the Bethe roots receive finite-size corrections, {\it i.e.}
\begin{align}
p \rightarrow p + g^{4}\delta p.
\end{align}
Consequently, the asymptotic energy ${\cal E}(p)$ also gets corrected
\begin{align}
\mathcal{E}(p) &= \sqrt{1+4 g^2\sin^2\frac{p + \delta p}{2}} =
\sqrt{1+4 g^2\sin^2\frac{p}{2}} + g^{6}\sin p\, \delta p  +
\mathcal{O}(g^{8}).
\end{align}
The correction $\delta p$ can be computed from the Bethe equations and knowledge of the S-matrix \cite{Bajnok:2009vm}. Define the following function
\begin{align}
BAE_k =-\left(\frac{u_k + i}{u_k-i}\right)^2\prod _{j=1}^{M} \frac{u_k-u_j + 2 i}{u_k-u_j-2 i}.
\end{align}
The correction to the momentum $\delta p$ is then described by the following set of equations
\begin{align}\label{eqn;deltaPviaPhi}
\sum_i \frac{\partial BAE_k}{\partial p_i}\delta p_i = \Phi_k, \qquad k=1,\ldots M,
\end{align}
where momentum and rapidity are related via $u = \cot\frac{p}{2}$ and $\Phi_k$ is given by \cite{deLeeuw:2010ed}
\begin{align}
\Phi_k = \sum_{Q=1}^{\infty}\int dv \frac{d\tilde{p}^Q}{dv} \frac{g^4}{(v^2+Q^2)^2}\mathrm{str}_Q\left[\pi_Q(g) \mathbb{S}_{Q1}(v,u_1)\ldots\partial_v\mathbb{S}_{Qk}(v,u_k)\ldots \mathbb{S}_{QM}(v,u_M)\right].
\end{align}
Explicitly, for the orbifolded Konishi state, we find 
\begin{align}
 \Phi_1 = -\Phi_2 = \sum_{Q=1}^{\infty}\int dv\left[\frac{\sqrt{3}-3 v}{Q^2-3 v^2}-\frac{2 (v+\frac{1}{\sqrt{3}})(Q^2+(v+\frac{1}{\sqrt{3}})^2+1)}{((Q-1)^2+(v+\frac{1}{\sqrt{3}})^2)((Q+1)^2+(v+\frac{1}{\sqrt{3}})^2
)}\right]\frac{Y_Q}{2}.
\end{align}
Carefully computing the NLO energy correction to the orbifolded Konishi state then results in
\begin{align}
 E_{NLO} = -g^6\left[\frac{1}{12}\sin^4\frac{\alpha}{2} +  \frac{3}{8}\sin^2\frac{\alpha}{2}\right]. 
\end{align}
The second term is recognizable as the leading order wrapping correction in $\beta$-deformed theory \cite{Arutyunov:2010gu,Beccaria:2010kd,deLeeuw:2010ed}.

\subsubsection*{TBA approach}

The computation of the NLO contribution in the TBA approach follows a similar path. The only difference is that the asymptotic value of the Bethe root $p_i$ receives a correction coming from the main TBA equation (\ref{eq:ExactBethe}) rather than relating it to the derivative of the S-matrix via $\Phi_k$. Let us schematically write (\ref{eq:ExactBethe}) as
\begin{align}
\pi (2n_k+1) = p_k J + \log S(u,-u) + \mathcal{R}_k.
\end{align}
Varying the above over the asymptotic solution we find
\begin{align}
0 = \delta p_k  J + \delta\mathcal{R}_k.
\end{align}
In order for this to agree with the solution of (\ref{eqn;deltaPviaPhi}), $\delta\mathcal{R}_k$ should be related to $\Phi_k$ as
\begin{align}
\Phi_k = -\delta\mathcal{R}_k.
\end{align}
For the Konishi operator this was indeed shown to be the case by relating $\delta\mathcal{R}_k$ to the XXX-spin chain \cite{Balog:2010xa,Balog:2010vf}. A similar treatment is possible here. 

The crucial observation is that we can explicitly construct Y-functions based on the XXX-spin chain that satisfy an equation which strongly resembles the relevant TBA equation but with a different source term. Due to the similarities between the TBA equations of the orbifolded model and the undeformed model, it is easily seen that $\delta\mathcal{R}_k$ will satisfy the equations discussed in \cite{Balog:2010xa,Balog:2010vf}, albeit with a different source term. 

To make contact with an XXX-like model, consider the following transfer matrix
\begin{align}
t_{M}(v) = (M+1)[3(v-u_1)(v-u_2)-M(M+2)],
\end{align}
which satisfies
\begin{align}
t_{M}(v+i)t_{M}(v-i)=t_{M+1}(v)t_{M-1}(v)+t_{0}(v+(M+1)i)t_{0}(v-(M+1)i).
\end{align}
It is readily seen that $t_M(v)$ has two real roots $r_{a,b}$. Upon taking $u_1=-u_2=\frac{1}{\sqrt{3}}$ these roots exactly coincide with the roots $r$ of $Y_{M|vw}$. It is also easily checked that $Y_{M|vw}$ to lowest order can be expressed via $t_{M}(v)$.

Moreover, we also find that $Y^{\circ}_{M+1}(v)$ can be written in terms of $t_M(v)$ in the following way
\begin{align}
&Y^{\circ}_{M}(v) = \left.\frac{g^4\sin^4(\alpha/2)}{(v^2+Q^2)^2}\frac{t_{M-1}(v)^2}{t^+_+ t^+_- t^-_+ t^-_-}\right|_{u_1=-u_2=\frac{1}{\sqrt{3}}}, && t^\pm_\pm = t_{0}(v\pm i(M\pm1)).
\end{align}
This puts us into exactly the same position as in \cite{Balog:2010xa,Balog:2010vf}. It is then straightforward to apply the therein obtained results to find 
\begin{align}
\delta\mathcal{R} = \frac{1}{2\pi}\sum_{Q=1}^{\infty} \int{ dv \, \partial_k Y^{\circ}_{Q}(v)}.
\end{align}
This agrees exactly with $\Phi$, proving that both ways of computing the NLO contribution (or factually the finite-size correction to the Bethe root) are in perfect agreement.

\subsubsection*{Ground state}

In case of the ground state, there is no momentum $p$ that gets corrected, nor are there Bethe roots. In other words, at next-to-leading-order, the only contribution comes from expanding $\tilde{p}^Q$ and $Y_Q$. These contributions give rise to
\begin{align}
E_{NLO} = \frac{16(J+1)}{\pi} g^{2J+2}\sin^4\frac{\alpha}{2}\sum_{Q=1}^{\infty}\int dv Q^2 \frac{Q^2-v^2}{(v^2+Q^2)^{J+2}} 
\end{align}
For $J>1$, this then becomes
\begin{align}
 E_{NLO}=  16g^{2J+2} \sin^4\frac{\alpha}{2} 
\frac{ \Gamma(J+\tfrac{1}{2})\zeta(2 J-1)}{\sqrt{\pi} \Gamma(J)}.
\end{align}
The contribution for $J=1$ again diverges, while this time the $J=2$ contribution is finite.

The above arguments can actually be extended all the way up to order $g^{2J}$. Due to the absence of excitations, the wrapping corrections up to double wrapping (order $g^{2J}$) are solely described by the expansion of $Y_Q$. In fact, since the transfer matrix is trivial, the wrapping correction to this order is simply obtained from the small $g$ expansion of
\begin{align}
-\frac{8\sin^4\frac{\alpha}{2}}{\pi}\sum_{Q=1}^{\infty} \int dv \frac{d\tilde{p}^Q}{dv} e^{-J\tilde{\mathcal{E}}_Q(v)} Q^2.
\end{align}
At order $g^{2J}$, however, $Y_Q$ will obtain corrections from the other Y-functions, specifically from $Y_{M|vw}$. It would be interesting to study this correction.

\section{Conclusions}

In this paper we have argued that the TBA equations for simple deformations of the $\ads$ string should be given by rather modest modifications of the TBA equations for the undeformed model. It would be interesting to see a first principles derivation of the general canonical TBA equations, to see the way in which these deformations enter explicitly, both for orbifolded and $\beta$-deformed theories. Specifically, when considering the orbifold deformation of this paper, the modifications disappear completely upon going to the simplified TBA equations, a fact which should be seen from the canonical TBA equations directly, if and when they are found. It would also be of interest to do a detailed analysis of the ground state solution for this orbifolded model, which is no longer protected from receiving finite size corrections, by considering the double wrapping correction to the energy. Knowing the simplified TBA equations for the ground state, we were able apply the contour deformation trick, combined with knowledge of the asymptotic solution to find a set of TBA equations that describes the orbifolded Konishi state. Both this set of equations and L\"usher's approach were then used to independently determine the NLO wrapping correction to the energy of this state, giving perfect agreement. An interesting topic to consider in this specific model would be the next-to-next-to-leading-order (NNLO) correction from both the TBA and L\"usher approaches, providing the next nontrivial test of their perturbative equivalence, as well as giving insights on higher order wrapping corrections.

\section*{Acknowledgments}

We are grateful to Gleb Arutyunov, Zoltan Bajnok and Tomasz Lukowski for useful discussions, and to Sergey Frolov for useful comments on the manuscript. The work by S.T. is part of the VICI grant 680-47-602 of the Netherlands Organization for Scientific Research (NWO) and the ERC Advanced Grant research programme No. 246974, {\it``Supersymmetry: a window to non-perturbative physics"}.

\appendix

\section{Twisted transfer matrix}
\label{ap:twistedtransfermatrix}
The eigenvalue of the twisted transfer matrix for an
anti-symmetric bound state representation with the bound state
number $Q$ is given by the following formula, generalizing the
result of \cite{Arutyunov:2009iq}
\begin{eqnarray}\label{eqn;FullEignvalue}
&&T_{Q,1}(v\,|\,\vec{u})=\prod_{i=1}^{K^{\rm{II}}}{\textstyle{\frac{y_i-x^-}
{y_i-x^+}\sqrt{\frac{x^+}{x^-}}
\, +}}\\
&&
{\textstyle{+}}\prod_{i=1}^{K^{\rm{II}}}{\textstyle{\frac{y_i-x^-}{y_i-x
^+}\sqrt{\frac{x^+}{x^-}}\left[
\frac{x^++\frac{1}{x^+}-y_i-\frac{1}{y_i}}{x^++\frac{1}{x^+}-y_i-\frac{1
}{y_i}-\frac{2i Q}{g}}\right]}}\prod_{i=1}^{K^{\rm{I}}}
{\textstyle{\left[\frac{(x^--x^-_i)(1-x^-
x^+_i)}{(x^+-x^-_i)(1-x^+
x^+_i)}\frac{x^+}{x^-}  \right]}}\nonumber\\
&&{\textstyle{+}}
\sum_{k=1}^{Q-1}\prod_{i=1}^{K^{\rm{II}}}{\textstyle{\frac{y_i-x^-}{y_i-
x^+}\sqrt{\frac{x^+}{x^-}}
\left[\frac{x^++\frac{1}{x^+}-y_i-\frac{1}{y_i}}{x^++\frac{1}{x^+}-y_i-\frac{1}{y_i}-\frac{2ik}{g}}\right]}}
\left\{\prod_{i=1}^{K^{\rm{I}}}{\textstyle{\lambda_+(v,u_i,k)+}}\right.\left.\prod_{i=1}^{K^{\rm{I}}}{\textstyle{\lambda_-(v,u_i,k)}}\right\}\nonumber\\
&&\quad -\sum_{k=0}^{Q-1}\prod_{i=1}^{K^{\rm{II}}}
{\textstyle{\frac{y_i-x^-}{y_i-x^+}\sqrt{\frac{x^+}{x^-}}\left[\frac{x^+
-\frac{1}{x^+}-y_i-\frac{1}{y_i}}
{x^+-\frac{1}{x^+}-y_i-\frac{1}{y_i}-\frac{2ik}{g}}\right]}}\prod_{i=1}^
{K^{\rm{I}}}{\textstyle{\frac{x^+-x^+_i}{x^+-x^-_i}\sqrt{\frac{x^-_i}{x^
+_i}} \left[1-\frac{\frac{2ik}{g}}{v-u_i+\frac{i}{g}(Q-1)
}\right]}}\times\nonumber\\
&&\quad\times
\left\{e^{i\alpha}\prod_{i=1}^{K^{\rm{III}}}{\textstyle{\frac{w_i-x^+-\frac{1}{x^+}
+\frac{i(2k-1)}{g}}{w_i-x^+-\frac{1}{x^+}+\frac{i(2k+1)}{g}}+ }}
e^{-i\alpha}\prod_{i=1}^{K^{\rm{II}}}{\textstyle{\frac{y_i+\frac{1}{y_i}-x^+-\frac
{1}{x^+}+\frac{2ik}{g}}{y_i+\frac{1}{y_i}-x^+-\frac{1}{x^+}+\frac{2i(k+1
)}{g}}}}\prod_{i=1}^{K^{\rm{III}}}{\textstyle{\frac{w_i-x^+-\frac{1}{x^+
}+\frac{i(2k+3)}{g}}{w_i-x^+-\frac{1}{x^+}+\frac{i(2k+1)}{g}}}}\right\}.
\nonumber
\end{eqnarray}
Here the twist $e^{i\alpha}$  enters only the last line.
Eigenvalues are parametrized by solutions of the auxiliary Bethe
equations:
\begin{eqnarray}
\label{bennote}
\prod_{i=1}^{K^{\rm{I}}}\frac{y_k-x^-_i}{y_k-x^+_i}\sqrt{\frac{x^+_i}{x^
-_i}}&=&e^{i\alpha}
\prod_{i=1}^{K^{\rm{III}}}\frac{w_i-y_k-\frac{1}{y_k}-\frac{i}{g}}{w_i-y
_k-\frac{1}{y_k}+\frac{i}{g}},\\
\prod_{i=1}^{K^{\rm{II}}}\frac{w_k-y_i-\frac{1}{y_i}+\frac{i}{g}}{w_k-y_
i-\frac{1}{y_i}-\frac{i}{g}} &=& e^{2i\alpha}\prod_{i=1,i\neq
k}^{K^{\rm{III}}}\frac{w_k-w_i+\frac{2i}{g}}{w_k-w_i-\frac{2i}{g}}.\nonumber
\end{eqnarray}
In the formulae above the variable
$$
v=x^++\frac{1}{x^+}-\frac{i}{g}Q=x^-+\frac{1}{x^-}+\frac{i}{g}Q\,
$$
takes values in the mirror theory rapidity plane, i.e. $x^\pm = x(v
\pm {i\ov g}Q)$ where $x(v)$ is the mirror theory $x$-function. As
was mentioned above, $u_j$ take values in string theory $u$-plane,
and therefore $x_j^\pm = x_s(u_j \pm {i\ov g})$ where $x_s(u)$ is
the string theory $x$-function. These two functions are given by
\begin{align}
&x(u) = \frac{1}{2}(u-i\sqrt{4-u^2}), && x_s(u) = \frac{u}{2}(1+\sqrt{1-\frac{4}{u^2}}).
\end{align}
Finally, the quantities {\small
$\lambda_{\pm}$ are
\begin{eqnarray}\nonumber \hspace{-1cm}
\lambda_\pm(v,u_i,k)&=&\frac{1}{2}\left[1-\frac{(x^-_ix^+-1)
  (x^+-x^+_i)}{(x^-_i-x^+)
  (x^+x^+_i-1)}+\frac{2ik}{g}\frac{x^+
  (x^-_i+x^+_i)}{(x^-_i-x^+)
  (x^+x^+_i-1)}\right.\\ \label{eqn;lambda-pm}
&&~~~~~~~~~~~~\left.\pm\frac{i x^+
  (x^-_i-x^+_i)}{(x^-_i-x^+)
 (x^+x^+_i-1)}\sqrt{4-\left(v-\frac{i(2k-Q)}{g}\right)^2}\right]\,
 .
\end{eqnarray}
}

The S-matrix in the string-mirror region $S_{\sl(2)}^{1_*Q}$ is
found in \cite{Arutyunov:2009kf} (see also \cite{Bajnok:2009vm}) and it has
the following weak-coupling expansion
$$
S_{\sl(2)}^{1_*Q}(u,v)=S_0(u,v)+g^2 S_{2}(u,v)+\ldots \, ,
$$
where {\small
\begin{align}
S_0(u,v) = -\frac{\big[(v-u)^2+(Q+1)^2\big]\big[Q-1 + i (v-u)
)\big]}{(u-i)^2 \big[Q-1-i( v-u)\big]}.
\end{align}
} and {\small
\begin{align}
S_2(u,v)& =-S_0(v,u)
\frac{2\big[2Q(u-i)+(u+i)(v^2+Q^2+2v(u-i))\big]}{(v^2+Q^2)
(1+u^2)}+\\
\nonumber &
\frac{S_0(v,u)}{1+u^2}\Big[4\gamma+\psi\left(1+\frac{Q+iv}{2}\right)+\psi\left(1-\frac{Q+iv}{2}\right)+\psi\left(1+\frac{Q-iv}{2}\right)+
\psi\left(1-\frac{Q-iv}{2}\right)\Big]\, .
\end{align}
}These expressions are enough to build up the two leading terms in
the weak-coupling expansion of the asymptotic function $Y^o_Q$.

\section{Critical behaviour of the asymptotic solution}
\label{ap:criticality}

There is a considerable number of additional roots that plays a role in the TBA for orbifolded Konishi as the coupling is increased, and their behaviour changes as the twist is varied. We will give an overview of this behaviour, but refrain from e.g. presenting the sets of driving terms these roots would result in, as their derivation is immediate once the roots are known. Let us start with the roots for $Y_{M|vw}$-functions.

\subsection{$Y_{M|vw}$-functions}

\begin{figure}
\begin{center}
 \includegraphics[width=3.5in]{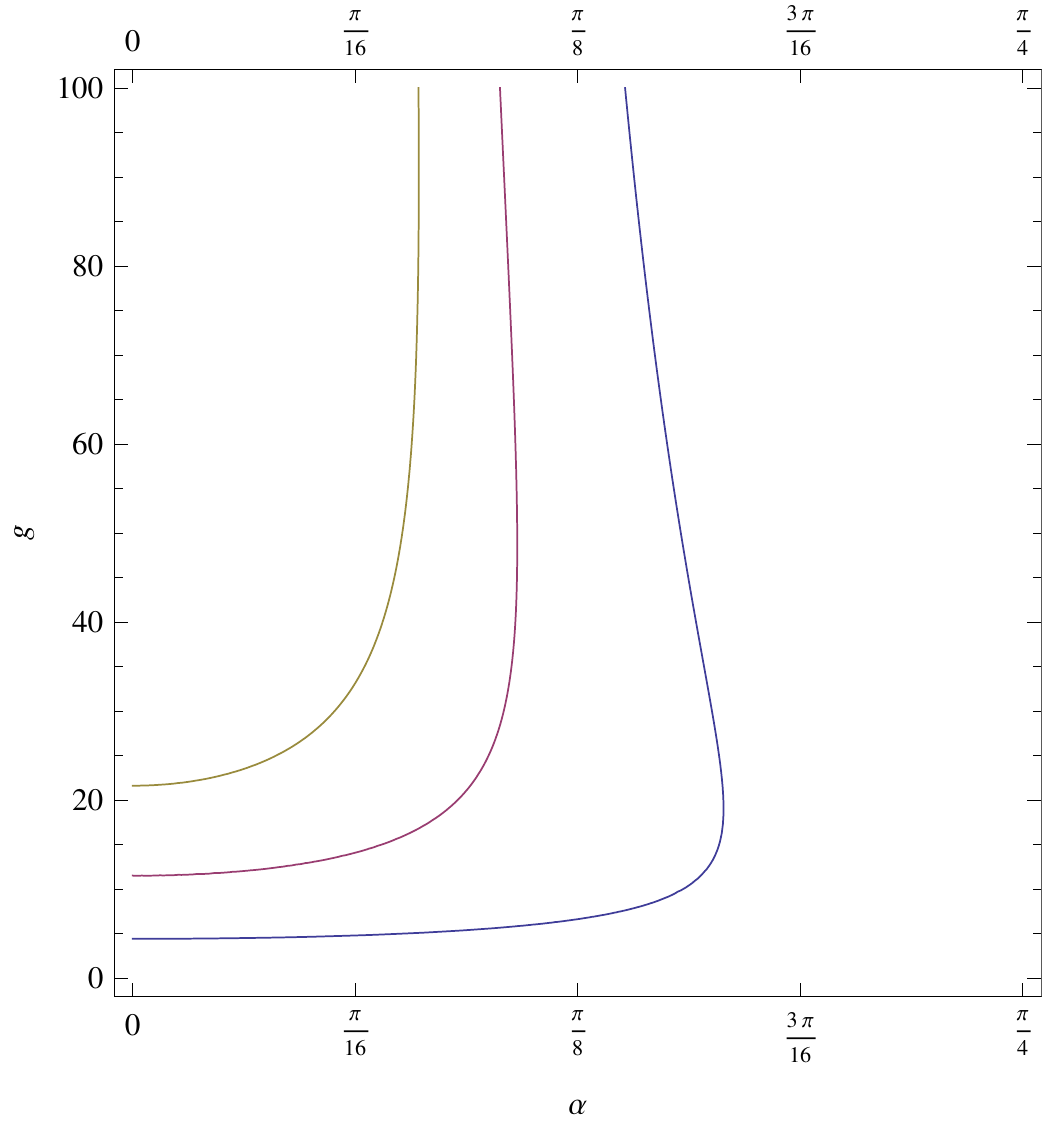}
\end{center}
\caption{The asymptotic critical values corresponding to $Y_{1|vw}$ (blue), $Y_{2|vw}$ (red), and $Y_{3|vw}$ (yellow).}
\label{fig:YMvwcrit}
\end{figure}

The critical behaviour of roots of $1+Y_{M|vw}$ is largely analogous to that observed for the asymptotic solution of the undeformed model \cite{Arutyunov:2009ax}, only there are twice the number of roots. In the undeformed case, $1+Y_{M|vw}$ has two roots that cross the integration contour at a certain value of the coupling, resulting in driving terms above the so called critical value, $g_{crit}$. These roots are imaginary and move towards the lines $\pm i/g$, where they each split in two roots lying symmetrically on the lines $i/g$ and $-i/g$. For the asymptotic solution, the first few critical values were found to be $4.429$, $11.512$ and $21.632$ corresponding to $Y_{1|vw}$, $Y_{2|vw}$ and $Y_{3|vw}$ respectively. For our orbifolded model, we find that $1+Y_{M|vw}$ in general has four extra roots, in addition to the noncritical ones mentioned in the main text. Two of these correspond directly to the roots present in the undeformed case, and we can plot their critical values as a function of twist. The result, immediately requiring some further discussion, is presented in figure \ref{fig:YMvwcrit}.
Note that these plots directly reproduce the critical values observed for the undeformed model. The main interesting feature is that the plot of the critical value as presented, is not really a function of the twist; it it multi-valued. However, looking at a specific twist, this simply corresponds to having two critical values. As the reader might have guessed by now, this second critical value simply corresponds to the critical value for the second pair of roots present in our orbifolded model. Beyond the point where the two critical values merge, the roots no longer play a role in the TBA; they do not cross the integration contour.

The interesting question that remains is whether the upper part of these curves actually touches the $g$-axis at a finite (but clearly large) value of the coupling, meaning that the second pair of roots should also play a role in the undeformed model, though only at very large coupling. This is in fact not the case, as we can show nicely that these curves should close in on the $g$-axis at infinite coupling; let us do so explicitly for the  curve corresponding to $Y_{1|vw}$.

The critical value curve for $Y_{1|vw}$ corresponds to the curve $T_{21}(\pm i/g)=0$, which follows by construction of the Y-functions, \textit{cf.} (\ref{eq:YmvwinT}). Considering that in the  large $g$ limit, the solution of the BY equation is given by \cite{Arutyunov:2009ax}
\begin{equation}
p = \sqrt{\tfrac{2\pi}{g}} - \tfrac{1}{g},
\end{equation}
the expansion of $T_{21}(\pm i/g)$ around $g = \infty$ gives
\begin{equation}
T_{21}(\pm i/g) = 4(1-\cos{\alpha}) + \sqrt{\mathcal{O}(\tfrac{1}{g})},
\end{equation}
with the first correction proportional to $c + \cos(\alpha)$\footnote{$T_{Q1}(\pm i/g)$ gives $2Q(1-\cos{\alpha})$, with corrections proportional to something of the form $\tilde{c} + \cos(\alpha)$.}, where $c$ is some constant. This shows that at zero twist, the second zero of $T_{21}(\pm i/g)$ lies at infinite coupling.

\subsection{$Y_{M|w}$-functions}

\begin{figure}
\begin{center}
 \includegraphics[width=3.5in]{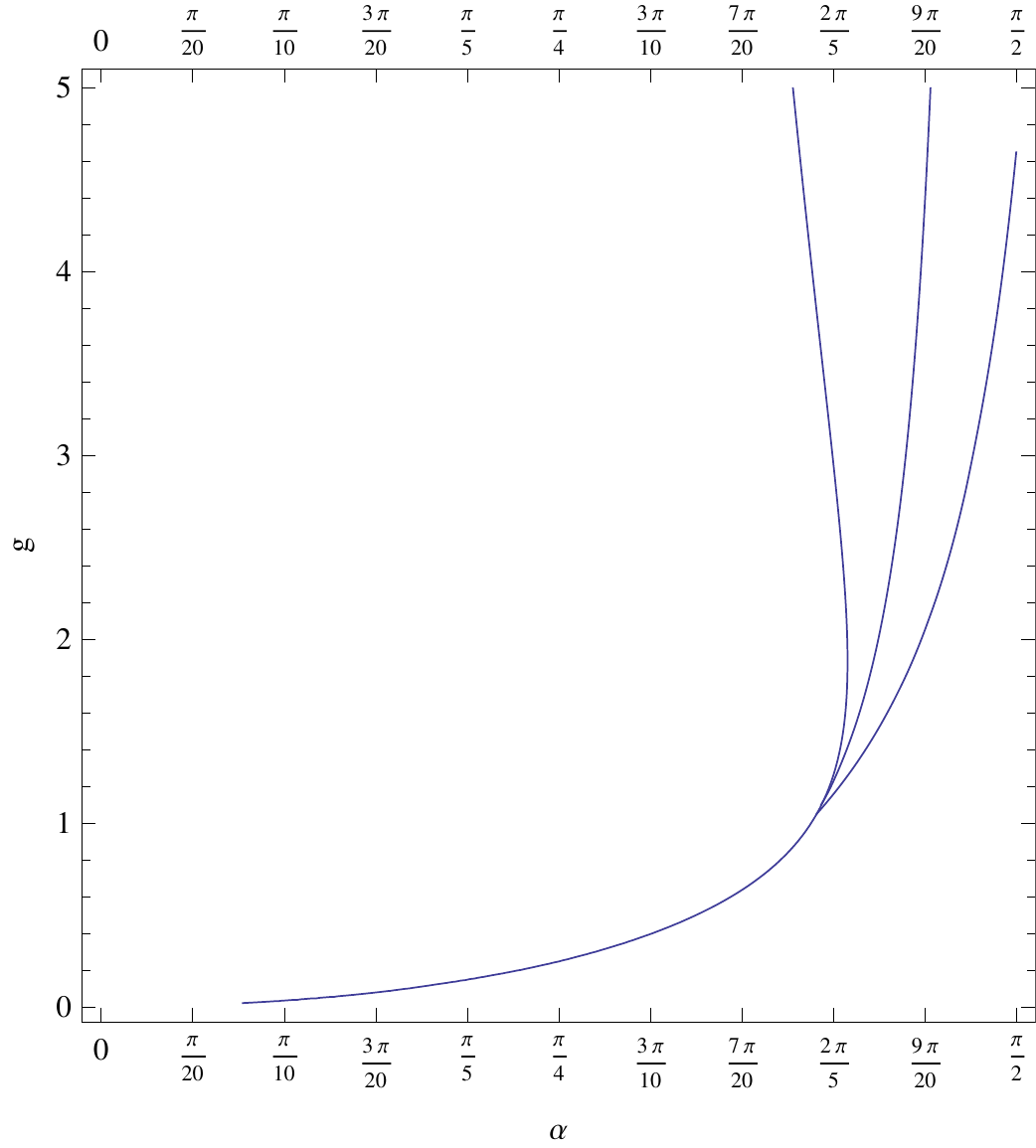}
\end{center}
\caption{The asymptotic critical values corresponding for $Y_-$. Note the possible existence of four critical values.}
\label{fig:YMwcrit1}
\end{figure}

While the analytic structure of the $Y_{M|vw}$-functions was still similar to the undeformed model, the case of $Y_{M|w}$-functions, where we include $Y_-$ as $-Y_{0|w}$, is qualitatively different and more involved. In general, these functions have four roots, and these four roots can show three different types of critical behaviour. The type of critical behaviour depends on the index $M$ of the $Y$ function, and the specific value of the twist, and can be quite intricate. Let us illustrate the general discussion with a plot of the critical values for $Y_-$ in figure \ref{fig:YMwcrit1}.
The way to interpret the above plot is to consider a specific twist, and look at the number of critical values. This can be one, two, three, or four, depending on the twist chosen. The rightmost curve corresponds to $T_{11}(-i/g)=0$, analogous to the above discussion for the $Y_{M|vw}$-functions. The following three types of critical behaviour can then occur:

\bigskip

\noindent \textbf{Type I:}

\noindent Provided we cross the critical value corresponding to the curve $T_{11}(-i/g)=0$ first (\textit{i.e.} $\alpha$ being smaller than the intersection point of the curves), there are simply two roots that enter the equations, and another two that enter after crossing the second critical value, just as for $Y_{M|vw}$.

\bigskip

\noindent \textbf{Type II:}

\noindent Beyond this point however, four roots will generically enter the equations after crossing the lowest critical value. A second set of roots will approach from outside the integration contour, and merge with the first set of roots exactly at the second critical value, pinching the integration contour. Immediately after, these merged roots will split in two again, where now half the contribution from all \emph{eight} roots needs to be taken into account, in accordance with the pinching.

Provided the twist is such that we can still cross the curve $T_{11}(-i/g)=0$, the first crossing corresponds to the second set of roots moving away again after merging at $\pm i/g$, leaving us with half the contribution from four roots, only to be completed again to half the contribution from all eight roots after crossing the curve a second time.

\bigskip

\noindent \textbf{Type III:}

\noindent As type II, but beyond the point where we can cross the curve $T_{11}(-i/g)=0$.

\bigskip

\begin{figure}[h]
\begin{center}
\includegraphics[width=4in]{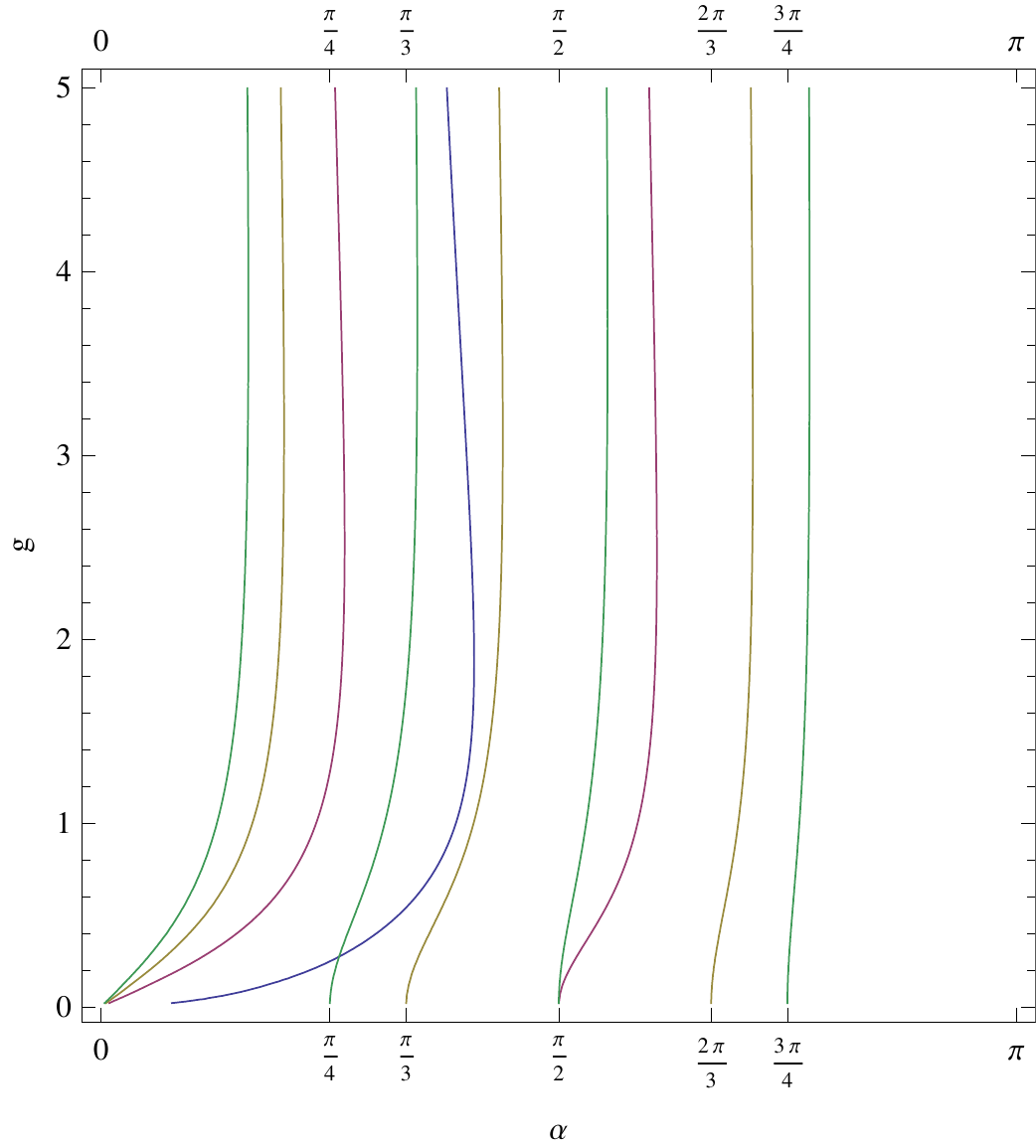}
\end{center}
\caption{The curves $T_{1j}(-i/g)=0$ for $j=1,2,3,4$, corresponding to critical values for $Y_-$(blue), $Y_{1|w}$(red), $Y_{2|w}$(yellow), and $Y_{3|w}$(green) respectively.}
\label{fig:YMwcrit2}
\end{figure}

Similar behaviour occurs for the $Y_{M|w}$-functions, with the change that there are multiple separate curves corresponding to the critical values; in general the critical value for a $Y_{M|w}$ function starts from zero $M+1$ times, at $\alpha = 0\mod\pi/(M+1)$. This has been illustrated in figure \ref{fig:YMwcrit2}. The complete picture of the asymptotic criticality is then obtained by adding additional curves to figure \ref{fig:YMwcrit2}, so that a picture qualitatively the same as figure \ref{fig:YMwcrit1} arises from each point where the critical value is zero. We refrain from presenting these curves here, as they provide no insight, and are time-consuming, though conceptually simple, to find.

Finally we would like to come back to the continuity of the hybrid equations, considered as a function of twist. While the relationship between $L_{\scriptscriptstyle TBA}$ and $J$ is discontinuous in the zero twist limit, the full equation, equation \ref{eq:hybridQ=1}, is of course continuous. The extra $2 \tH$ appears as the contribution from the extra roots of $1 - Y_-$ discussed above. In the zero twist limit, these roots move towards $\pm \infty -i/g$, while their critical value goes to zero simultaneously, \textit{cf.} figure \ref{fig:YMwcrit1}. Analyzing the resulting driving terms, the only term that survives in the limit of roots with large real parts is the term $\tfrac{1}{2}\log{\tfrac{S_{yQ}(r^-,v)S_{yQ}(-r^-,v)}{S_{yQ}(r^+,v)S_{yQ}(-r^+,v)}}$, where $r$ is the real part of any of the four roots. Taking the limit $r\rightarrow \infty$ in the expression for this kernel gives exactly $ -2 i \log{\tfrac{x^+}{x^-}} = 2 \tH$.

\bibliographystyle{JHEP}
\bibliography{OrbifoldKonishi}

\end{document}